\documentclass[english]{article}
\usepackage[latin9]{inputenc}
\usepackage{float}
\usepackage{amsmath}
\usepackage{subfig}
\usepackage{graphicx}

\makeatletter

\floatstyle{ruled}
\newfloat{algorithm}{tbp}{loa}
\providecommand{\algorithmname}{Algorithm}
\floatname{algorithm}{\protect\algorithmname}

\usepackage{babel}
\begin{document}

\title{On the CAD-compatible conversion of S-patches}

\author{Péter Salvi\\
Budapest University of Technology and Economics\\
}
\maketitle
\begin{abstract}
S-patches have many nice mathematical properties. It is known since
their first appearance, that any regular S-patch can be exactly converted
into a trimmed rational Bézier surface. This is a big advantage compared
to other multi-sided surface representations that have to be approximated
for exporting them into CAD/CAM systems. The actual conversion process,
however, remained at a theoretical level, with bits and pieces scattered
in multiple publications. In this paper we review the entirety of
the algorithm, and investigate it from a practical aspect.
\end{abstract}

\section{\label{sec:Introduction}Introduction}

S-patches offer a mathematically perfect generalization of Bézier
surfaces to any number of sides. The three-sided S-patch \textendash{}
the Bézier triangle \textendash{} is widely adopted, but there is
yet no standard representation for surfaces with more than four sides.

The lack of success for S-patches is likely due to their complex control
net structure, and the large number of control points. These, as it
turns out, can be (partially) solved by the automatic generation of
the control network using only boundary constraints~\cite{Salvi:2019:KEPAF}.

Another common issue with $n$-sided surface representations is the
conversion to tensor product patches, which is crucial for processing
models in CAD/CAM systems. Usually some fitting method is used, which
either creates $n$ quadrilaterals sharing a central vertex~\cite{Piegl:1999},
or a larger four-sided region with approximated trimming curves. The
former does not ensure $C^{\infty}$ continuity inside the patch,
while the latter has an inherent asymmetry, as well as parameterization
issues~\cite{Vaitkus:2018}.

Here S-patches seem to have an advantage: it is shown in the original
paper~\cite{Loop:1989} that any regular S-patch can be exactly converted
into a trimmed rational Bézier surface. The algorithm for this has
not been described in detail, however, so in our article we will go
through all the required steps.

In Section~\ref{sec:Previous-Work} we will briefly review the related
publications. In Section~\ref{sec:S-patches} we will define the
S-patch both in itself, and as the composition of Bézier simplexes.
Then the conversion process is described in Section~\ref{sec:Conversion-to-quadrilateral}.
An example and the discussion of some practical issues conclude the
paper.

\section{\label{sec:Previous-Work}Previous Work}

S-patches were first published by Loop and DeRose~\cite{Loop:1989}
in 1989, but the techniques needed for the conversion process were
developed earlier: see Ramshaw's wonderful book~\cite{Ramshaw:1987}
on blossoming, and DeRose's paper~\cite{DeRose:1988} on the composition
of Bézier simplexes. A more efficient variation~\cite{DeRose:1993}
of the latter was developed a few years later.

In the following sections, we will try to connect the dots, and insert
a few missing pieces (notably the polarization of Wachspress coordinates
in Section~\ref{subsec:Polarization}), hopefully making tensor product
conversion easy to understand and implement.

\section{\label{sec:S-patches}S-patches}

An $n$-sided S-patch\footnote{This paper deals only with \emph{regular} S-patches, as only these
have the required properties for conversion.} is defined over a regular $n$-gon, parameterized by Wachspress barycentric
coordinates $\boldsymbol{\lambda}=(\lambda_{1},\dots,\lambda_{n})$~\cite{Hormann:2006}.
Its control points $\{P_{\mathbf{s}}\}$ are labeled by $n$ non-negative
integers whose sum is the \emph{depth}\footnote{Sometimes also referred to as the \emph{degree}, since the boundaries
of an S-patch are Bézier curves of degree $d$.} of the surface ($d$). We will use the notation $L_{n,d}$ for the
set of all such labels, and $(s_{1},\dots,s_{n})=\mathbf{s}\in L_{n,d}$
for one particular label. The norm of a label is its sum, i.e., $|\mathbf{s}|=d$.

The surface point corresponding to a domain point with barycentric
coordinates $\boldsymbol{\lambda}$ is defined as
\begin{equation}
S(\boldsymbol{\lambda})=\sum_{\mathbf{s}\in L_{n,d}}P_{\mathbf{s}}\cdot B_{\mathbf{s}}^{d}(\boldsymbol{\lambda})=\sum_{\mathbf{s}\in L_{n,d}}P_{\mathbf{s}}\cdot\binom{d}{\mathbf{s}}\cdot\prod_{i=1}^{n}\lambda_{i}^{s_{i}},\label{eq:spatch}
\end{equation}
where $B_{\mathbf{s}}^{d}(\boldsymbol{\lambda})$ are Bernstein polynomials,
and
\begin{equation}
\binom{d}{\mathbf{s}}=\frac{d!}{\prod_{i=1}^{n}s_{i}!}
\end{equation}
are their multinomial coefficients.

\subsection{S-patches as Bézier simplexes}

An $(n-1)$-dimensional simplex has $n$ vertices $V_{i}$ (e.g.~in
2D a triangle, in 3D a tetrahedron). Any point can be uniquely expressed
by the affine combination of these vertices:
\begin{equation}
p=\sum_{i=1}^{n}\lambda_{i}V_{i},\quad\sum_{i=1}^{n}\lambda_{i}=1.
\end{equation}
The coefficients $\lambda_{i}$ are called the barycentric coordinates
of $p$ relative to the simplex.

A Bézier simplex of dimension $n-1$ and degree $d$ is a polynomial
mapping from the barycentric coordinates relative to an $(n-1)$-dimensional
simplex, given in Bernstein form:
\begin{equation}
\sum_{\mathbf{s}\in L_{n,d}}P_{\mathbf{s}}\cdot B_{\mathbf{s}}^{d}(\boldsymbol{\lambda}),
\end{equation}
where $P_{\mathbf{s}}$ are called the control points of the Bézier
simplex. We can see that this is the same as Eq.~(\ref{eq:spatch}),
i.e., the S-patch is a Bézier simplex that maps from an $(n-1)$-dimensional
simplex to 3D, restricted to the embedded domain polygon (the $n$
generalized barycentric coordinates).

Let us introduce the following notations for a Bézier simplex $S$:
\begin{itemize}
\item $n_{S}$ : \# of arguments (i.e., domain simplex dimension + 1)
\item $d_{S}$ : degree
\item $\delta_{S}$ : dimension of the control points
\end{itemize}
These form the characteristic triple $\phi(S)=(n_{S},d_{S},\delta_{S})$.
For example, for an $n$-sided S-patch of depth $d$, $\phi(S)=(n,d,3)$.

\section{\label{sec:Conversion-to-quadrilateral}Conversion to trimmed Bézier
patches}

According to the original paper~\cite{Loop:1989}, any $n$-sided S-patch
of depth $d$ can be converted into an $m$-sided rational S-patch
of depth $(n-2)d$. Also, a four-sided S-patch of depth $d$ can be
converted into a tensor product Bézier patch of degree $d$.

Consequently, the conversion is done in two stages. First the $n$-sided
S-patch is replaced with a four-sided one (Section~\ref{subsec:Conversion-to-quadrilateral}),
then the tensor product form is computed (Section~\ref{subsec:Conversion-to-tensor}),
resulting in a surface of degree $(n-2)d$.

\subsection{\label{subsec:Conversion-to-quadrilateral}Conversion to quadrilateral
S-patches}

The Wachspress coordinates of an $n$-sided regular polygon have the
form
\begin{equation}
\lambda_{i}(p)=\frac{\prod_{j\neq i-1,i}D_{j}(p)}{\sum_{k=1}^{n}\prod_{j\neq k-1,k}D_{j}(p)},\label{eq:wachspress-regular}
\end{equation}
where $D_{j}(p)$ is the signed distance of $p$ from the $j$-th
side.

When $p$ is given by barycentric coordinates, this can also be expressed
as a rational Bézier simplex $W_{n}$, that maps from a 2D simplex
to a simplex of dimension $n-1$. This is not an obvious result, since
Eq.~(\ref{eq:wachspress-regular}) involves Euclidean distances.
For the exact construction, and the handling of rational simplexes,
see Section~\ref{subsec:Polarization}.

It is also shown that $W_{n}$ is \emph{pseudoaffine}, i.e., it has
an affine left inverse $W_{n}^{-1}$. With this, we can write
\begin{equation}
S\circ W_{n}=S\circ W_{n}\circ(W_{4}^{-1}\circ W_{4})=(S\circ W_{n}\circ W_{4}^{-1})\circ W_{4}.
\end{equation}
In other words, the four-sided version of the $n$-sided depth $d$
S-patch $S$ is the composition of three Bézier simplexes, with the
following characteristic triples: $\phi(W_{4}^{-1})=(4,1,3)$, $\phi(W_{n})=(3,n-2,n)$,
and $\phi(S)=(n,d,4)$.

Let us see how to generate control points for each of these Bézier
simplexes. Since we will have to work with barycentric coordinates
relative to some simplex, define the \emph{canonical simplex} as the
canonical basis and the origin.\emph{ }In 2D, this will be the triangle
$(1,0)$, $(0,1)$, $(0,0)$, so the barycentric coordinates for a
$(u,v)$ point will be $(u,v,1-u-v)$.

The Bézier simplex $W_{4}^{-1}$ defines the vertices of the four-sided
domain in the plane. Since the tensor product patch will be defined
on $[0,1]\times[0,1]$, we will use the same square as the image of
the 3D simplex. Thus the control points of $W_{4}^{-1}$ are:
\begin{align}
P_{1000} & =(0,1,0), & P_{0100} & =(1,1,-1),\nonumber \\
P_{0010} & =(1,0,0), & P_{0001} & =(0,0,1).
\end{align}

Getting the control points of $W_{n}$ needs a bit more work, see
the next section.

The control points of $S$ are just the control points of the original
$n$-sided S-patch. The astute reader must have noticed that $\delta_{S}=4$.
This is because $W_{n}$ is rational, and the composition algorithm
needs $S$ to be also in homogenized barycentric form (see below).

\subsubsection{\label{subsec:Polarization}Polarization of the Wachspress coordinates}

Homogeneous coordinates are normally represented by adding an extra
``weight'' coordinate, so $(x,y,z)\equiv(wx,wy,wz,w)$, and projection
works by dividing with the weight. With Bézier simplexes, we use another
type of homogenization, that allows us to use the simplex composition
algorithm as it is, even for the rational case: we take the barycentric
coordinates (relative to some simplex) and use them as ``normal''
coordinates in a higher dimension.

If we use the canonical simplex, an $(x,y,z)$ point has the barycentric
coordinates $(x,y,z,1-x-y-z)$, so in general the homogenized form
will be $(x,y,z)\equiv(wx,wy,wz,w(1-x-y-z))$. Projection is done
by dividing with the sum of all coordinates.

The homogenized form of Wachspress coordinates is thus $\left\{ \prod_{j\neq i-1,i}D_{j}(p)\right\} $.
We will work out the Bézier simplex coordinates for this by using
its \emph{blossom} or \emph{polarization}. For any homogeneous polynomial
$Q(u)$ of degree $d$, there is a symmetric multilinear function
$q$ of $d$ arguments that agrees with it on its diagonal:
\begin{align}
q(u_{1},\dots,u_{d}) & =q(u_{\pi_{1}},\dots,u_{\pi_{d}}),\label{eq:symmetry}\\
q(u_{1},\dots,\alpha u_{k_{1}}+\beta u_{k_{2}},\dots,u_{d}) & =\alpha q(u_{1},\dots,u_{k_{1}},\dots,u_{d})\nonumber \\
 & +\beta q(u_{1},\dots,u_{k_{2}},\dots,u_{d}),\label{eq:multilinearity}\\
q(u,\dots,u) & =Q(u).\label{eq:agreement}
\end{align}
Eq.~(\ref{eq:symmetry}) shows symmetry for permutations, Eq.~(\ref{eq:multilinearity})
multilinearity, and Eq.~(\ref{eq:agreement}) the diagonal agreement.

The rational Bézier simplex control points for $Q$ are easily computed
from its polarization:
\begin{equation}
P_{\mathbf{s}}^{Q}=q(\underbrace{V_{1},\dots,V_{1}}_{s_{1}},\underbrace{V_{2},\dots,V_{2}}_{s_{2}},\dots,\underbrace{V_{n},\dots,V_{n}}_{s_{n}}),\label{eq:blossom}
\end{equation}
where $V_{i}$ are the vertices of the domain simplex.

The blossom of Wachspress coordinates for a regular $n$-gon is given
as
\begin{equation}
q(p_{1},\dots,p_{n-2})_{i}=\frac{1}{(n-2)!}\cdot\sum_{\pi\in\Pi(n-2)}\prod_{\substack{k=1\\
j\neq i-1,i
}
}^{n-2}D_{j}(p_{\pi_{k}}),
\end{equation}
where $\Pi(n-2)$ is the set of permutations of $\{1,\dots,n-2\}$,
and in the product $k$ runs from $1$ to $n-2$ while $j$ goes from
$1$ to $n$ skipping $i-1$ and $i$. Now Eq.~(\ref{eq:blossom})
can be used to compute the coordinates of $W_{n}$.

The exact position and rotation of the $n$-sided polygon affect the
quality of the generated control net. A good choice is to use cyclic
polygons on the circle that has its origin at $(0.5,0.5)$, and a
radius of $0.5$.

\subsubsection{Simplex composition}

Now that all three of the Bézier simplexes are well-defined, we can
turn our attention to their composition. Let $F$ and $G$ denote
Bézier simplexes, and $H=F\circ G$. Then
\begin{equation}
\phi(H)=\left(n_{G},d_{F}\cdot d_{G},\delta_{F}\right),
\end{equation}
and the control points are given by
\begin{equation}
P_{\mathbf{s}}^{H}=\sum_{\substack{|\mathbf{s}_{1}|+|\mathbf{s}_{2}|+\dots+|\mathbf{s}_{d_{F}}|=d_{G}\\
\mathbf{s}_{1}+\mathbf{s}_{2}+\dots+\mathbf{s}_{d_{F}}=\mathbf{s}
}
}\frac{\binom{d_{G}}{\mathbf{s}_{1}}\cdots\binom{d_{G}}{\mathbf{s}_{d_{F}}}}{\binom{d_{F}\cdot d_{G}}{\mathbf{s}}}\Delta_{\mathbf{0}_{n_{F}}}(P_{\mathbf{s}_{1}}^{G},\dots,P_{\mathbf{s}_{d_{F}}}^{G}),\label{eq:comp-cp}
\end{equation}
where $\mathbf{0}_{n_{F}}$ is a multi-index or vector of length $n_{F}$
consisting entirely of zeros, and
\begin{equation}
\Delta_{\mathbf{s}}(p_{1},\dots,p_{k})=\begin{cases}
\sum_{i=1}^{n_{F}}p_{1}^{i}P_{\mathbf{s}+\mathbf{e}_{i}}^{F} & \text{if }k=1,\\
\sum_{i=1}^{n_{F}}p_{k}^{i}\Delta_{\mathbf{s}+\mathbf{e}_{i}}(p_{1},\dots,p_{k-1}), & \text{otherwise.}
\end{cases}
\end{equation}
Here $p_{k}^{i}$ is the $i$-th coordinate of $p_{k}$, and $\mathbf{s}+\mathbf{e}_{i}$
is the multi-index $\mathbf{s}$ with the $i$-th position increased
by one.

\subsubsection{\label{subsec:Efficient-composition-algorithm}Efficient composition
algorithm}

The equations shown in the previous section, when implemented naïvely,
are highly inefficient. DeRose et al.~\cite{DeRose:1993}~show an
algorithm for computing it in a less computationally intensive manner.
It is based on caching the results of the recursive calls to $\Delta$,
and using the fact that $\Delta$ is symmetric in its arguments.

\begin{algorithm}
\textsc{compose}($F$, $G$):\\
\phantom{xx}$P_{\mathbf{s}}^{H}\leftarrow\mathbf{0}_{\delta_{H}}$
for all $\mathbf{s}\in L_{n_{H},d_{H}}$\\
\phantom{xx}$P_{\mathbf{s}}^{\hat{F},1}\leftarrow P_{\mathbf{s}}^{F}$
for all $\mathbf{s}\in L_{n_{F},d_{F}}$\\
\phantom{xx}\textbf{fn} \textsc{blossom}($k$, $p$):\\
\phantom{xx}\phantom{xx}\textbf{for} $\mathbf{s}$ in $L_{n_{F},d_{F}-k}$:\\
\phantom{xx}\phantom{xx}\phantom{xx}$P_{\mathbf{s}}^{\hat{F},k+1}\leftarrow\sum_{i=1}^{n_{F}}p^{i}P_{\mathbf{s}+\mathbf{e}_{i}}^{\hat{F},k}$\\
\phantom{xx}\textbf{fn} \textsc{rec}($k$, $\mathbf{s}_{\min}$,
$\mathbf{s}_{\Sigma}$, $c$, $\mu$):\\
\phantom{xx}\phantom{xx}\textbf{if} $k=d_{F}$:\\
\phantom{xx}\phantom{xx}\phantom{xx}$P_{\mathbf{s}_{\Sigma}}^{H}\leftarrow P_{\mathbf{s}_{\Sigma}}^{H}+c\cdot P_{\mathbf{0}_{n_{F}}}^{\hat{F},d_{F}+1}$\\
\phantom{xx}\phantom{xx}\textbf{else}:\\
\phantom{xx}\phantom{xx}\phantom{xx}$\mathbf{s}\leftarrow\mathbf{s}_{\min}$\\
\phantom{xx}\phantom{xx}\phantom{xx}\textbf{while} $\mathbf{s}\neq\emptyset$:\\
\phantom{xx}\phantom{xx}\phantom{xx}\phantom{xx}\textsc{blossom}($k+1$,
$P_{\mathbf{s}}^{G}$)\\
\phantom{xx}\phantom{xx}\phantom{xx}\phantom{xx}\textsc{rec}($k+1$,
$\mathbf{s}$, $\mathbf{s}_{\Sigma}+\mathbf{s}$, $c\cdot\binom{d_{G}}{\mathbf{s}}/\mu$,
$\mu+1$)\\
\phantom{xx}\phantom{xx}\phantom{xx}\phantom{xx}$\mu\leftarrow1$\\
\phantom{xx}\phantom{xx}\phantom{xx}\phantom{xx}$\mathbf{s}\leftarrow$\textsc{next}($\mathbf{s}$)\\
\phantom{xx}\textsc{rec}($0$, $\mathbf{0}_{n_{G}}+\mathbf{e}_{n_{G}}\cdot d_{G}$,
$\mathbf{0}_{n_{G}}$, $d_{F}!$, $1$)\\
\phantom{xx}$P_{\mathbf{s}}^{H}\leftarrow P_{\mathbf{s}}^{H}/\binom{d_{H}}{\mathbf{s}}$
for all $s\in L_{n_{H},d_{H}}$\\
\phantom{xx}\textbf{return} $P^{H}$

\caption{\label{alg:Efficient-composition}Efficient composition of two Bézier
simplexes.}

\end{algorithm}
The pseudocode is shown in Algorithm~\ref{alg:Efficient-composition}.
Here \textsc{next}($\mathbf{s}$) is the next multi-index in a lexicographical
ordering, e.g. $0021\rightarrow0030\rightarrow0102\rightarrow0111\rightarrow\dots$,
returning\emph{ $\emptyset$} after the last element ($3000$ in the
example). Since $\Delta$ is now computed only for this lexicographical
permutation of its arguments, its result is multiplied by $d_{F}!$
divided by the factorials of the argument multiplicities (built up
by the variable $\mu$).

The above paper suggests the use of arrays for the control points,
indexed by the lexicographical order of the multi-index labels. A
conceptually simpler alternative is the use of a dictionary data structure
(a hash table) that maps control point positions to their labels.
Our experiments show that this is actually even faster, as the hash
function is cheaper to compute than the lexicographical index.

\subsubsection{Change of coordinates}

The homogenized control points computed above have the form $(wx,wy,wz,w(1-x-y-z))$.
To convert these to the usual homogeneous coordinates $(wx,wy,wz,w)$,
we just need to replace the last coordinate by the sum of all coordinates.

\subsection{\label{subsec:Conversion-to-tensor}Conversion to tensor product
form}

A (rational or non-rational) quadrilateral S-patch of depth $d$ has
the tensor product Bézier patch form
\begin{equation}
\hat{S}(u,v)=\sum_{i=0}^{d}\sum_{j=0}^{d}C_{ij}B_{i}^{d}(u)B_{j}^{d}(v)
\end{equation}
with the control points
\begin{equation}
C_{ij}=\sum_{\substack{\mathbf{s}\\
s_{2}+s_{3}=i\\
s_{3}+s_{4}=j
}
}\frac{\binom{d}{\mathbf{s}}}{\binom{d}{i}\binom{d}{j}}P_{\mathbf{s}}.
\end{equation}

\section{\label{sec:Case-Studies}Example}

\begin{figure}
\subfloat[S-patch control network]{\begin{centering}
\includegraphics[width=0.5\textwidth]{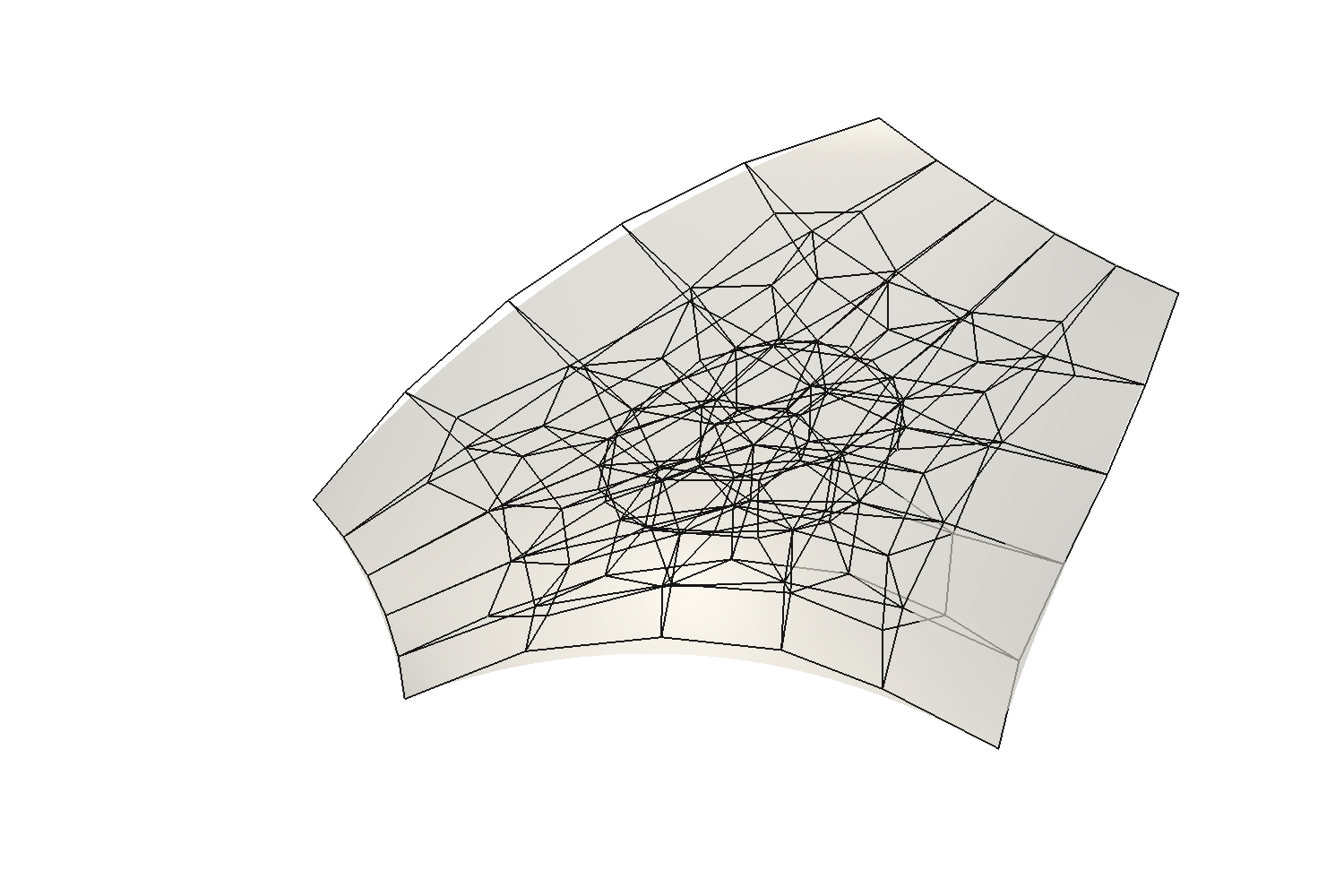}
\par\end{centering}
}
\subfloat[Contouring]{\begin{centering}
\includegraphics[width=0.5\textwidth]{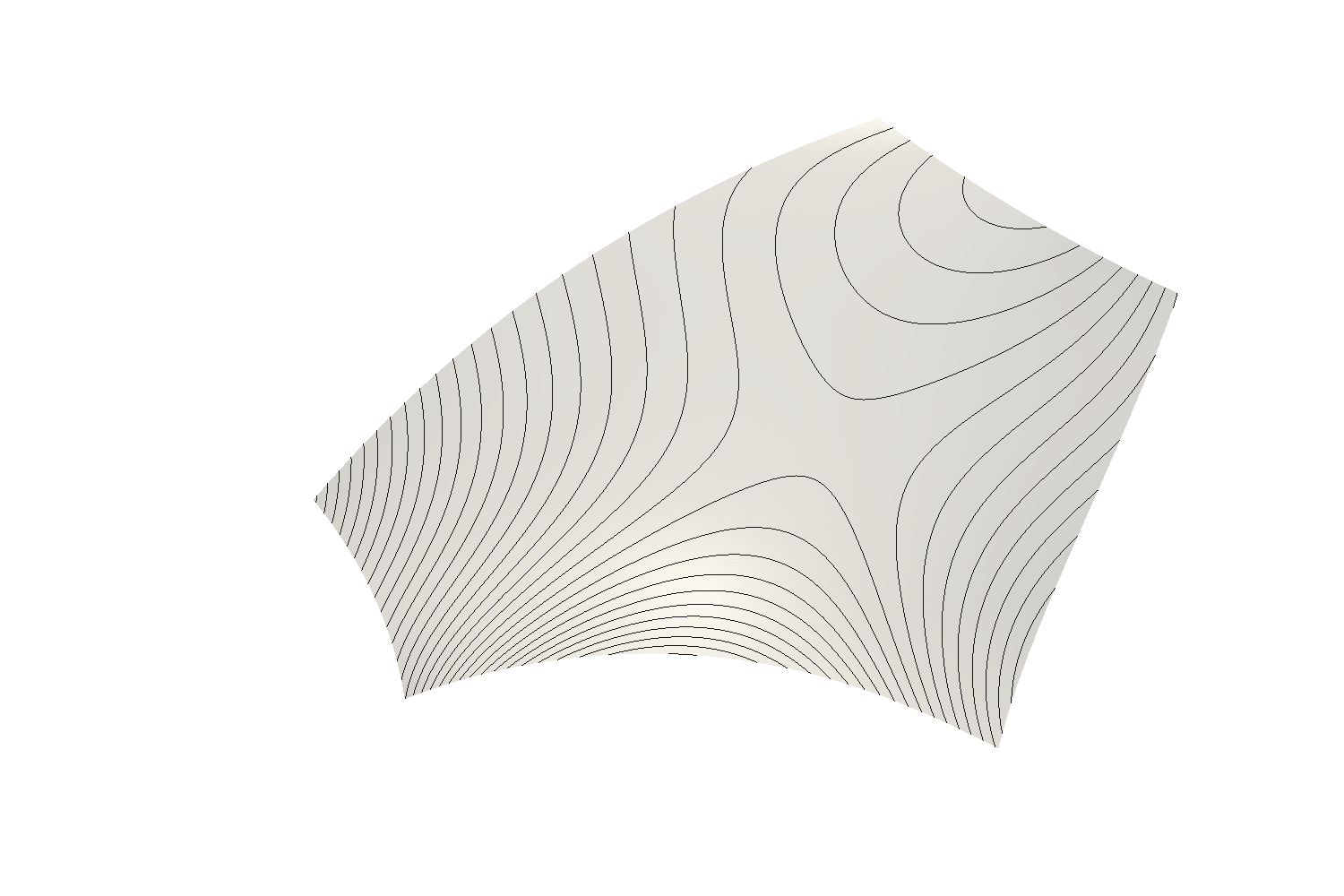}
\par\end{centering}
}
\par
\subfloat[Trimmed patch with control net]{\begin{centering}
\includegraphics[width=0.5\textwidth]{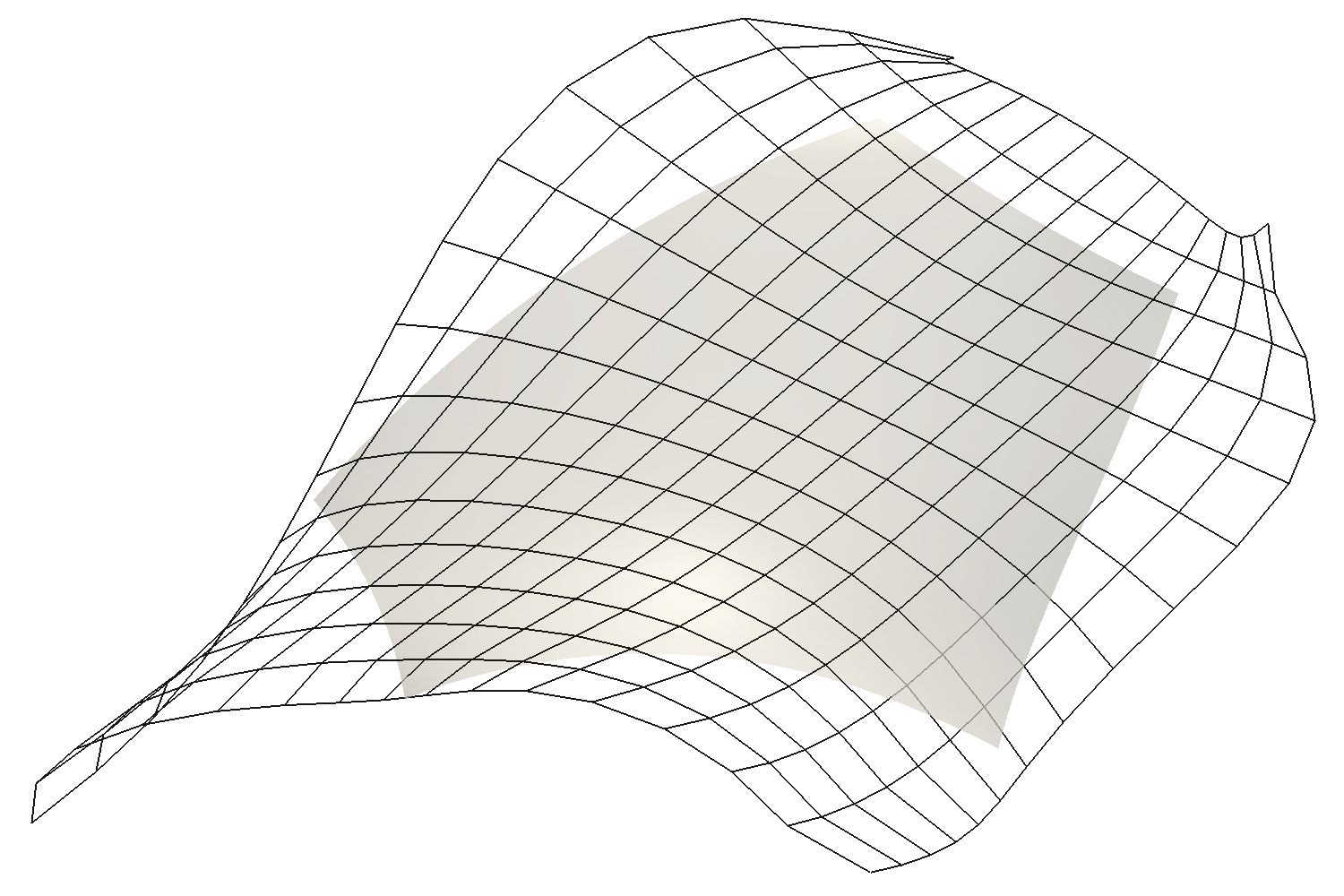}
\par\end{centering}
}
\subfloat[Full tensor product patch]{\begin{centering}
\includegraphics[width=0.5\textwidth]{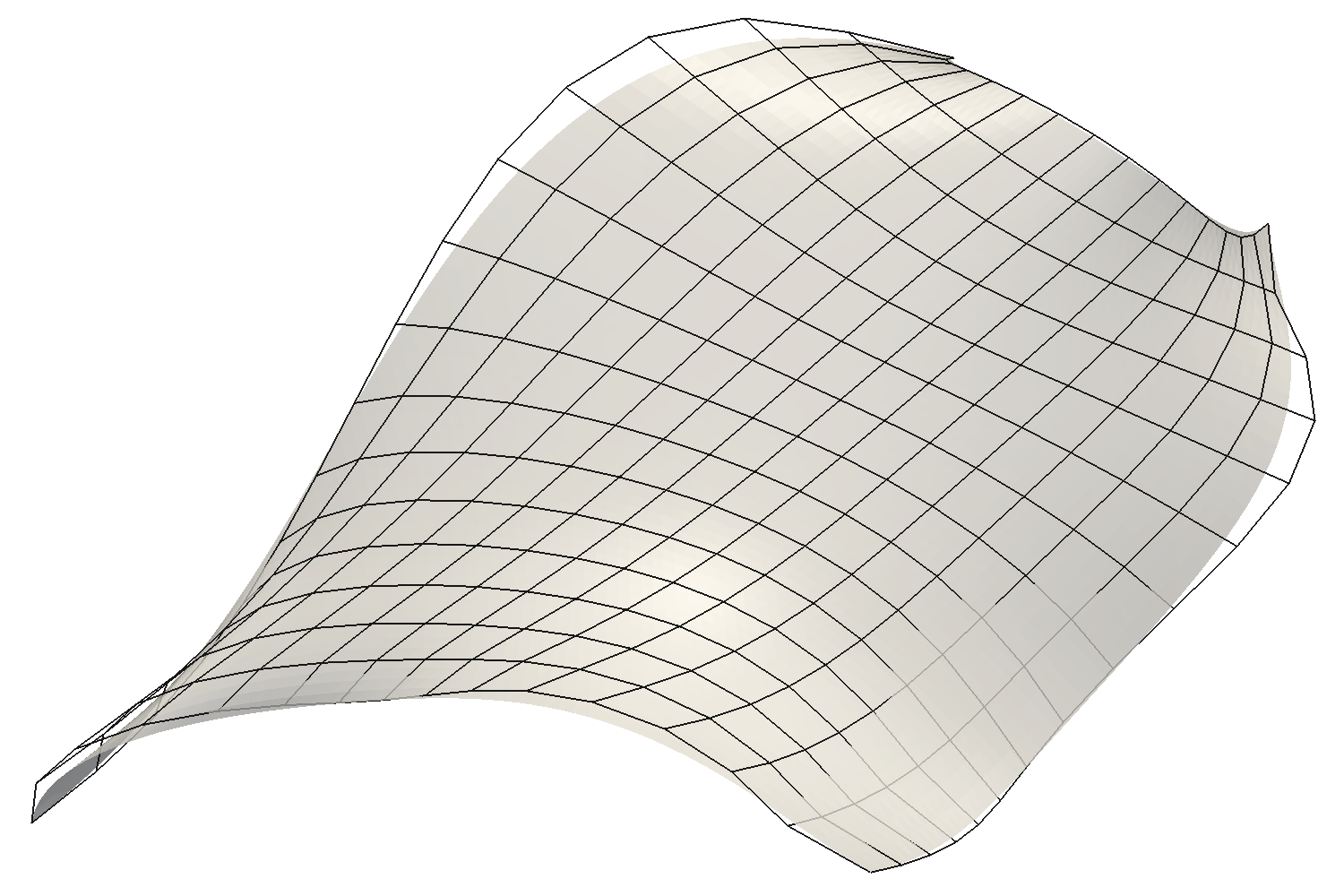}
\par\end{centering}
}
\caption{\label{fig:A-5-sided-S-patch}A 5-sided S-patch of depth 5.}
\end{figure}

Figure~\ref{fig:A-5-sided-S-patch} shows a 5-sided S-patch of depth
5, converted into a $15\times15$-degree tensor product rational Bézier
patch. The example model was generated by the ribbon-based algorithm
of the author~\cite{Salvi:2019:KEPAF}.

\section{Discussion}

We have seen how S-patches are transformed into trimmed Bézier patches.
Let us now look at some practical issues.

\subsection{Efficiency}

The algorithm in Section~\ref{subsec:Efficient-composition-algorithm}
is still computationally demanding. Converting a 5-sided S-patch of
depth 8 took more than 5 minutes on a 2.8GHz processor. There is a
much faster approach \textendash{} see the upcoming paper of the author~\cite{Salvi:2020}.

\subsection{Triangles}

Three-sided S-patches, i.e., Bézier triangles, are converted into
simple polynomial patches, since computing the barycentric coordinates
do not involve rational polynomials. Note however, that there are
alternative methods for the quadrilateral transformation of a triangular
patch, see e.g.~Warren's domain deformation method~\cite{Warren:1992}.

\subsection{Control net quality}

One issue with the conversion is that the quadrilateral control grid
may have outlier control points or spikes near the corners. This is
because the denominator of Wachspress coordinates vanish on a circle
around the domain, and these singularities undermine the stability
of these areas.

\section*{Conclusion}

The S-patch representation of multi-sided surfaces could be an important
asset in a modeling toolbox. It can be used to fill holes or create
smooth vertex blends, and then it can be exported to CAD/CAM systems
as a trimmed Bézier patch, without losing precision, thus making it
possible to create perfectly watertight models. In this paper we have
reviewed the steps required for the quadrilateral conversion, and
discussed some related questions.

\section*{Acknowledgements}

This work was supported by the Hungarian Scientific Research Fund
(OTKA, No.\ 124727). The author thanks Tamás Várady for his valuable
comments.

\bibliographystyle{plain}
\bibliography{cikkek,sajat}

\end{document}